# A Novel Distributed PV Power Forecasting Approach Based on Time-LLM

Huapeng Lin, Miao Yu[*]

*College of Electrical Engineering, Zhejiang University, Hangzhou, 310027, China*

[*]*Corresponding authors:* zjuyumiao@zju.edu.cn *(Miao Yu)*

**Abstract:** Distributed photovoltaic (DPV) systems are essential for advancing renewable energy applications and achieving energy independence. Accurate DPV power forecasting can optimize power system planning and scheduling while significantly reducing energy loss, thus enhancing overall system efficiency and reliability. However, solar energy's intermittent nature and DPV systems' spatial distribution create significant forecasting challenges. Traditional methods often rely on costly external data, such as numerical weather prediction (NWP) and satellite images, which are difficult to scale for smaller DPV systems. To tackle this issue, this study has introduced an advanced large language model (LLM)-based time series forecasting framework Time-LLM to improve the DPV power forecasting accuracy and generalization ability. By reprogramming, the framework aligns historical power data with natural language modalities, facilitating efficient modeling of time-series data. Then Qwen2.5-3B model is integrated as the backbone LLM to process input data by leveraging its pattern recognition and inference abilities, achieving a balance between efficiency and performance. Finally, by using a flatten and linear projection layer, the LLM's high-dimensional output is transformed into the final forecasts. Experimental results indicate that Time-LLM outperforms leading recent advanced time series forecasting models, such as Transformer-based methods and MLP-based models, achieving superior accuracy in both short-term and long-term forecasting. Time-LLM also demonstrates exceptional adaptability in few-shot and zero-shot learning scenarios. To the best of the authors' knowledge, this study is the first attempt to explore the application of LLMs to DPV power forecasting, which can offer a scalable, data-efficient solution that eliminates reliance on costly external data sources and improve real-world forecasting accuracy and feasibility.

1. Introduction

As a widely popular green energy source, solar energy has characteristics such as abundant resources, easy accessibility, and no pollution [1]. The International Energy Agency proposed a plan in its 2021 report to accelerate the development of renewable energy, particularly solar and wind power, aiming to achieve net-zero carbon emissions in global electricity supply by 2035 [2]. Based on their characteristics, solar power plants can be divided into centralized photovoltaic (PV) plants and distributed PV (DPV) plants. Centralized PV plants are usually built in regions with abundant solar radiation, offering the advantages of large-scale operation and centralized management, making them suitable for providing stable power output to the grid. DPV plants, on the other hand, are deployed mostly on the user side, such as residential rooftops, commercial and industrial buildings, and agricultural facilities. They are characterized by low cost, on-plant power generation, and on-plant consumption, making full use of dispersed resources to achieve energy self-sufficiency. This model not only reduces transmission losses but also alleviates the grid's load to a certain extent, offering vast development potential. However, due to the influence of solar radiation and environmental factors, the output power of grid-connected DPV systems exhibits intermittent and fluctuating characteristics. This poses significant challenges to the operation, scheduling, and planning of power systems. Moreover, accurate DPV power forecasting can help reduce energy losses, effectively improving solar energy utilization efficiency and enhancing the profitability of power stations [3]. Thus, there is an urgent demand for the accurate DPV power forecasting.

In recent years, the rapid development of deep learning technology has provided significant opportunities for PV power forecasting [4]. As an extension of artificial neural networks, deep learning builds multilayer neural networks to efficiently model complex nonlinear time series data. This technology significantly improves the adaptability of models while reducing dependence on complex feature engineering, effectively overcoming the limitations of traditional machine learning methods in nonlinear modeling [5]. Compared to traditional methods, deep learning shows significant advantages in capturing complex dynamic variations and long-term trends in PV power generation, becoming the mainstream method in PV power forecasting.

Sky images are attracting growing attention in PV forecasting as a direct data source reflecting the environmental features of PV power generation. Liu et al. [6] introduced a deep decomposition framework leveraging sky images for ultra-short-term PV power forecasting. This method utilizes image decomposition techniques combined with deep learning models to separate sky images into cloud motion features and clear sky backgrounds, effectively capturing dynamic cloud characteristics and significantly improving forecasting accuracy. Nonetheless, this technique depends on high-quality sky images and exhibits limited robustness under cloudy and challenging weather conditions. Zang et al. [7] developed a novel sky image-based framework that incorporates time-series image processing techniques and deep learning models to further enhance the ability to capture cloud changes under complex weather conditions. This method significantly improves forecasting stability under complex weather conditions, but its dependence on high-quality images and sophisticated equipment makes it challenging to apply to DPV power plants with low investment costs and limited hardware resources.

In addition to sky images, satellite images, as a data source with broader coverage, have also shown great potential in PV power forecasting. Wang et al. [8] introduced an ultra-short-term forecasting approach for DPV power based on satellite cloud images and long short-term memory (LSTM) networks. This approach improves forecasting accuracy significantly by combining occlusion features from satellite cloud images with meteorological and solar irradiance data from DPV plants. Research shows that incorporating cloud image data reduces the forecasting error from 16.15% to 2.73%, improving accuracy by 13.42%. Its strength lies in integrating cloud image features with time-series data to accurately model PV power changes in complex weather, reducing reliance on the high-precision meteorological data required by traditional models. Cheng et al. [9] proposed a method for short-term solar power forecasting by directly learning from satellite images and designed a region of interest selection technique to effectively reduce computational costs. The approach offers a broad forecasting range and high accuracy but faces challenges in modeling surface characteristics, particularly in high-resolution demand contexts. Wang et al. [10] developed a cross-modal deep learning method that combines satellite images with meteorological data, significantly enhancing the accuracy and spatiotemporal consistency of PV power forecasting. The advantage of this method lies in cross-modal data fusion learning, which effectively models the complex relationships between different data sources, thereby accurately capturing the dynamic characteristics of PV power and demonstrating greater adaptability and predictive capability.

Furthermore, numerical weather prediction (NWP) data, which offers long-term trend insights, has been extensively employed in PV power forecasting. Hu et al. [11] employed an improved PV power forecasting model combining LSTM networks and a self-attention mechanism, incorporating NWP for multi-step forecasting. This method leverages the long-term trend information provided by NWP data, significantly enhancing the model's adaptability to seasonal and climate changes, but it still falls short in capturing rapidly changing cloud dynamics. Mayer et al. [12] compared the performance of two NWP models, the global ECMWF and the regional AROME, in solar irradiance and PV power forecasting. Research indicates that ECMWF delivers superior accuracy in global forecasting, though its effectiveness lessens in regional PV forecasting scenarios. In contrast, the AROME model, with its higher spatial resolution, has an advantage in capturing regional variations, but its performance is limited in accurately locating convective cells due to the "double penalty" issue. Su et al. [13] proposed a short-term PV power forecasting method based on NWP correction, with a specific focus on the coupling relationships between meteorological variables. The method significantly enhances forecasting accuracy and robustness in cloudy weather by implementing bias corrections on traditional NWP models. Its advantage lies in integrating the synergy of meteorological variables, further improving forecasting accuracy. Nonetheless, the approach's dependence on high-quality historical data, alongside the high acquisition cost and limited resolution of NWP data, combined with the geographical dispersion of DPV plants, poses challenges for its application in distributed scenarios.

Moreover, researchers are actively investigating deep learning frameworks and signal decomposition methodologies by using historical data. Luo and Zhang [14] introduced a cascaded deep learning framework for PV power forecasting, which captures multi-level features of time series by integrating multi-resolution input data, thereby improving robustness and accuracy, particularly demonstrating strong adaptability across multi-scale time ranges. The flexibility of this framework allows it to integrate multiple data sources and adapt to complex real-world application scenarios. Wang et al. [15] developed a PV power forecasting interval method based on frequency domain decomposition and LSTM networks, which extracts periodic and non-periodic features of time series through frequency domain decomposition, significantly improving forecasting accuracy without losing global trends. This method excels in achieving both high forecasting accuracy and uncertainty quantification, playing a crucial role in ensuring grid scheduling safety and stability. Lin et al. [16] proposed a short-term DPV power forecasting model based on signal decomposition and temporal self-attention mechanisms. Using complete ensemble empirical mode decomposition for signal decomposition, it extracts key features and employs temporal self-attention to capture complex time dependencies. Experimental results indicate that this method outperforms traditional models in terms of forecasting accuracy and stability, offering an efficient and reliable solution for DPV power forecasting. Cao et al. [17] investigated a multi-time-scale PV power forecasting approach using an improved Stacking ensemble algorithm. By integrating various machine learning models, it achieved high accuracy and stability in both short- and medium-term forecasts. This approach's strength lies in leveraging the Stacking framework to seamlessly combine the predictive strengths of diverse models, performing exceptionally across varying time scales. Tao et al. [18] developed a Transformer-based daily forecasting model, combining data augmentation and PV physical modeling to enhance forecasting accuracy. The model extracts temporal and inter-feature dependencies in parallel, with data augmentation significantly boosting predictive performance. It surpasses existing long-sequence forecasting models across multiple public datasets, with its innovation focusing on segmenting and embedding methods to optimize input data semantics, catering to real-world grid scheduling requirements. Abdel-Basset et al. [19] proposed a novel deep learning architecture, PV-Net, which redesigns GRU units and incorporates residual connections to enhance the capture of spatiotemporal features. The model performs excellently in short-term PV power forecasting, particularly demonstrating significant advantages in handling non-stationary data and high-frequency variations.

Although the above methods have achieved significant progress, some pressing challenges still need to be addressed. First, some existing methods rely on high-quality external data, such as NWP and satellite images, which are costly to obtain and difficult to scale in DPV power plants. Meanwhile, to better model time series features, signal decomposition methods are widely used in PV power forecasting. By separating trends, periods, and residual, signal decomposition helps models capture sequence characteristics. However, some decomposition methods may inadvertently use future data from validation or test datasets to enhance smoothness, causing information leakage and exaggerating model performance [20,21]. Therefore, methods that rely solely on historical data for direct modeling become particularly important. This method prevents information leakage while efficiently capturing global trends and local dynamic characteristics.

In recent years, large language model (LLM) has achieved significant progress in natural language processing. These models possess strong reasoning capabilities, efficiently handling complex patterns while capturing long-term dependencies and deep associations between sequences. Leveraging multi-task pretraining, LLMs perform exceptionally well across diverse scenarios, especially showing robust adaptability under data scarcity and pronounced distribution changes. Building on these strengths, Time-LLM extends LLMs to time series forecasting by reprogramming input data to align with the language modality. Through prompt engineering and patch embedding, it effectively captures short-term dynamics and long-term dependencies in time series data. The framework uses a frozen pre-trained backbone model, avoiding the need for extensive fine-tuning. This significantly lowers training costs while maintaining strong generalization in zero-shot and few-shot learning scenarios, enabling state-of-the-art performance across diverse forecasting tasks.

So far, LLMs have demonstrated significant potential in load forecasting and wind power forecasting by leveraging their cross-modal learning and feature extraction capabilities. Lee and Rew [22] used BERT to transform numerical load data into prompts, enabling LLMs to capture complex dependencies and generalize with minimal data, reducing reliance on historical records. Their few-shot learning capability improved adaptability across different load patterns. Hu et al. [23] applied LLM transfer learning for load forecasting, maintaining high forecasting accuracy with limited training data and demonstrating its efficiency in minimizing data requirements. Prompt engineering further enhanced model interpretability and computational efficiency. Lai et al. [24] developed BERT4ST, where LLMs' self-attention captured spatiotemporal dependencies in wind power forecasting, improving the modeling of wind speed fluctuations and power output. Prompt engineering integrated domain knowledge, and multi-stage fine-tuning ensured stability across different datasets. However, despite their success in other fields, there still remains a gap in the application of LLMs to DPV power forecasting to the best of authors' knowledge. To fill in this gap, we first study how to exploit LLMs to improve DPV power forecasting by leveraging their pattern recognition and inference abilities. Based on the characteristics of LLMs, we improved and applied the Qwen2.5-3B [25] model as the backbone under the Time-LLM [26] framework, as shown in Fig. 1, conducting in-depth research and validation on complex time series data in DPV scenarios. The experiments encompass tasks including short-term and long-term forecasting, few-shot learning, and zero-shot learning, clearly showcasing the model's efficiency and robustness in DPV power forecasting. The main contributions are as follows:

(1) This study is the first attempt to explore the application of LLMs to DPV power forecasting, demonstrating their potential in this domain and enhancing the model's practicality by utilizing historical data, which eliminates reliance on costly external data sources and improves real-world applicability in DPV power plant scenarios.

(2) The Qwen2.5-3B model is integrated as the backbone of Time-LLM, leveraging its knowledge density and task adaptability to maintain computational efficiency while meeting the performance requirements of time series forecasting tasks.

(3) Systematic experimental validation of the improved Time-LLM model is conducted across various scenarios, including short-term forecasting, long-term forecasting, few-shot learning, and zero-shot learning. Results show that the improved framework outperforms most existing advanced time series forecast methods in these tasks.

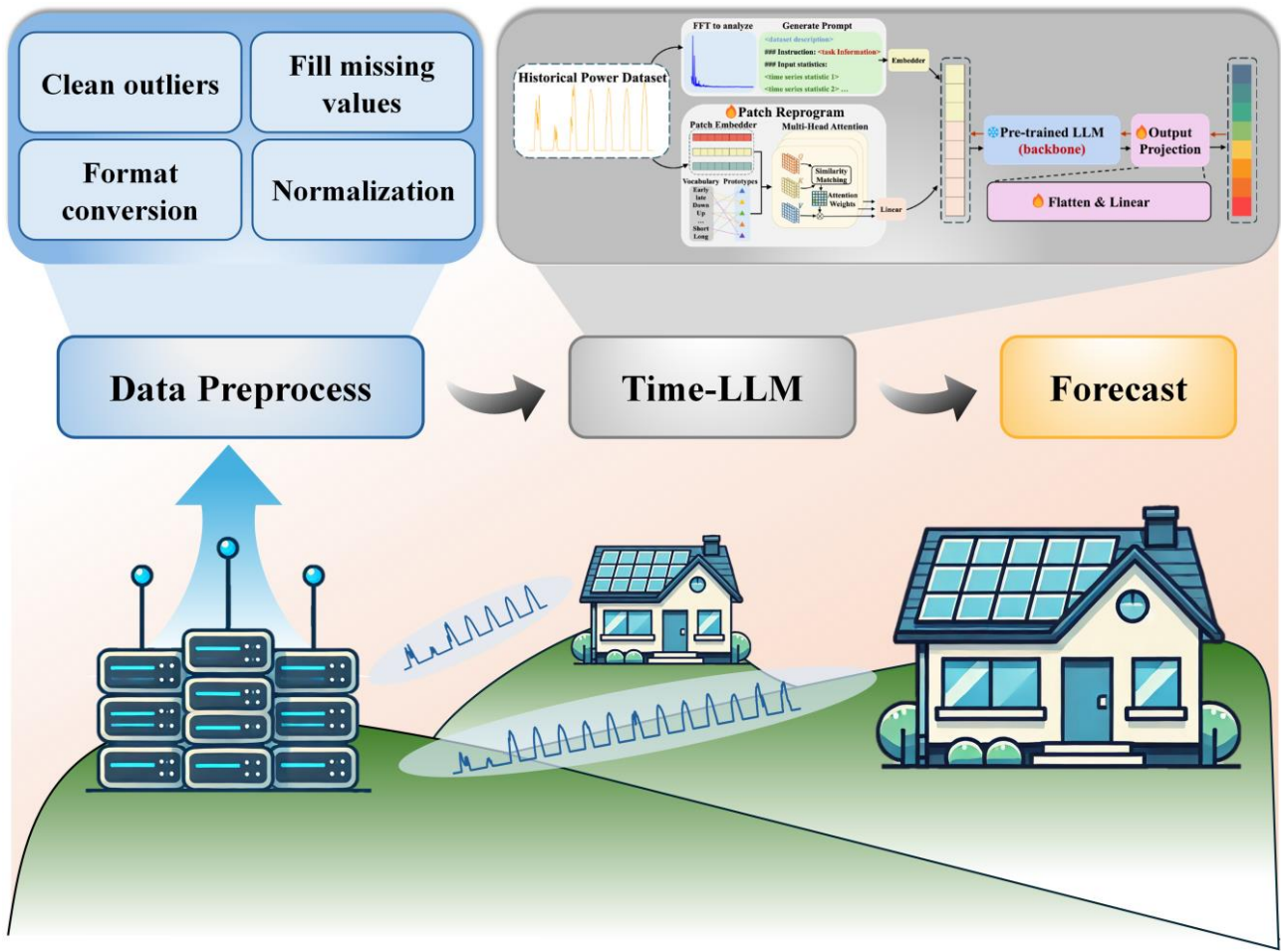

**Fig. 1 Our Framework for DPV Power Forecasting.** This framework consists of three main stages: (1) Data preprocess, including outlier cleaning, missing value filling, format conversion, and normalization of raw power data to ensure data consistency and reliability. (2) Time-LLM model, where the historical power datasets are reprogrammed and combined with prompt prefixes to perform time series modeling using a pre-trained LLM. (3) Forecast, generating power forecasting

results for DPV systems while enabling model adaptation and transferability across different DPV plants through zero-shot or few-shot learning.

## 2. TIME-LLM -Based Forecasting Method

Time-LLM employs reprogramming techniques to convert numerical time series into a format resembling natural language, thereby utilizing existing LLMs for forecasting without retraining LLMs. This approach not only significantly reduces computation resource requirement but also enhances generalization to diverse data. The core workflow begins with extracting periodic features using Fast Fourier Transform (FFT) and optimizing inputs with prompt engineering. The patch reprogramming and alignment are applied, allowing a pre-trained LLM to process the data and generate accurate forecasting results through output projection. Adopting a modular architecture, Time-LLM keeps the pre-trained LLM fixed to leverage its language understanding and reasoning abilities while avoiding costly retraining. Only the trainable components are limited to lightweight modules, including a linear mapping layer, a multi-head attention layer, and an output projection layer. The linear mapping layer aligns numerical data with the LLM's semantic space, and the multi-head attention mechanism enhances the interaction between numerical features and textual representations. The output projection layer maps the LLM's high-dimensional outputs to the target prediction space. For specific tasks, the model needs to tune hyperparameters such as patch length, sliding stride, and the number of attention heads. The detailed model structure is shown in Fig. 2.

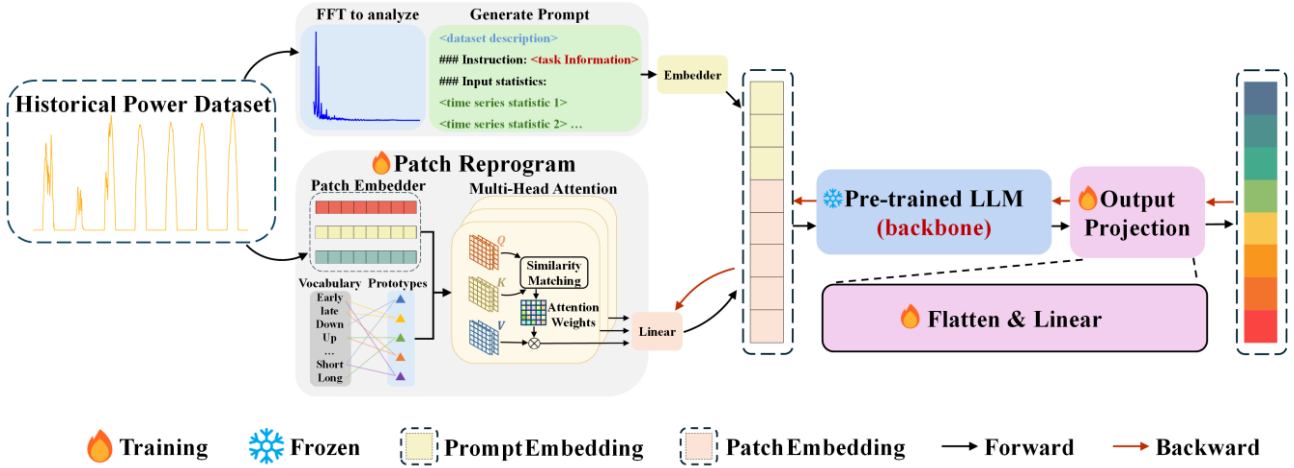

**Fig. 2 Time-LLM Model Structure**. The Time-LLM model structure leverages Fast Fourier Transform (FFT) to extract statistical information from time series to assist in generating prompt prefixes and aligns time series with the language modality through a reprogramming module. The aligned data is input into a frozen pre-trained LLM with the final forecasts generated through output projection.

### 2.1. Prompt-as-Prefix

Time-LLM first utilizes FFT to analyze the periodicity and trends within input data. For a time series $X = x_1, x_2, \ldots, x_L$,

where $L$ denotes the sequence length, the Discrete Fourier Transform (DFT) is mathematically defined as:

$$\hat{X}(f) = \sum_{q=1}^{Q} x_q \cdot e^{-j2\pi fq/Q} \tag{3-1}$$

where $\hat{X}(f)$ represents the signal in the frequency domain at frequency index $f$, encoding both the amplitude and phase of the frequency component. $x_q$ denotes the value of the time series at time $q$, and $f$ is the frequency index corresponding to a specific frequency component in the DFT. The total number of data points in the time series is $Q$, and $j$ is the imaginary unit, satisfying $j^2 = -1$.

The FFT is utilized to efficiently compute the DFT, revealing the periodic structure of the data. Analyzing the spectrum $\hat{X}(f)$ allows for the identification of key periodic components and data trends. The dataset context provides essential background information about the input time series, emphasizing its domain-specific characteristics. Task instructions guide the LLM in adapting patch embeddings for specific tasks. Statistical features such as minimum, maximum, trends, and lags are incorporated into the prompt text $P$, which combines dataset basics and task details to enhance pattern recognition and reasoning. A prompt example is shown in Fig. 3.

> The distributed photovoltaic dataset includes output power measurements from distributed photovoltaic systems located in ...
> Below is the information about the input time series:
> [Instruction]: forecast the next <Horizon> steps given the previous <Input Size> steps information;
> [Statistics]: The input has a minimum of <min_val>, a maximum of <max_val>, and a median of <median_val>. The overall trend is <upward or downward>. The top five lags are <lag_val>.

**Fig. 3 Prompt example.** ◇ and ◇ are task-specific configurations and calculated input statistics.

## 2.2. Input Embed

At the same time, Time-LLM partitions the long time series $X$ into multiple shorter patches. The sequence is patched as

$$S = S_1, S_2, \cdots, S_k \tag{3-2}$$

where $k$ is the number of patches and each patch $S_i$ consists of $m$ consecutive data points. Overlaps are allowed between adjacent patches. Specifically, the $i$-th patch can be defined as

$$S_i = x_{(i-1)s_d+1}, x_{(i-1)s_d+2}, \ldots, x_{(i-1)s_d+m} \tag{3-3}$$

where $s_d$ is the sliding step size (stride) that determines the overlap. When $s_d < m$, adjacent patches overlap, with $m - s_d$ data points shared between consecutive patches.

Using a linear embedding layer $f_{embed}$, each patch is mapped into a high-dimensional feature space to produce an embedding vector $e_i$ as

$$e_i = f_{embed}(S_i) \tag{3-4}$$

where $i$ represents $i$-th patch.

### 2.3. Patch Reprogram

Specifically, $W_{vocab}$ is the token embedding matrix from the pre-trained LLM. Time-LLM obtain it by loading the LLM's embedding parameters from the published checkpoint. Then we apply a linear mapping layer to project the embeddings into a prototype vector space, enabling alignment between these textual prototypes and the time series patches. The prototype vectors produced by the mapping layer are integrated with textual prototypes $T = Linear_{mapping}(W_{vocab})$, laying the foundation for subsequent alignment processes.

Subsequently, multi-head attention is utilized to reprogram the embeddings of time series and align them with textual prototypes. Embedding vectors $e_i$ serve as inputs for the query vectors, while textual prototypes $T$ are used to generate the key and value vectors. Specifically, the query $q_i^h$, key $k^h$, and value $v^h$ are computed using the following formulas:

$$q_i^h = W_q^h e_i, \quad k^h = W_k^h T, \quad v^h = W_v^h T \tag{3-5}$$

where $W_q^h$, $W_k^h$, $W_v^h$ denote the weight matrices for the $h$-th attention head.

Multi-head attention enhances expressive capacity by performing parallel computations of the weighted features across $H$ attention heads. For each attention head, the similarity between the query vector and the key vector is computed via dot product and normalized using the softmax to produce attention weights

$$\alpha_i^h = \text{softmax}\left(\frac{q_i^h \cdot k^h}{\sqrt{d_h}}\right) \tag{3-6}$$

where $\alpha_i^h$ denotes the attention distribution of the $h$-th head across each vocabulary prototype, and $d_h$ represents the dimensionality of each attention head. Subsequently, the value vectors are weighted by the attention scores to produce the context embeddings:

$$o_i^h = \alpha_i^h \cdot v^h \tag{3-7}$$

where $o_i^h$ denotes the output of the $h$-th attention head. All outputs from the multi-head attention mechanism are concatenated and passed through a linear projection layer $W_o$ to generate the final aligned embedding.

$$e_i' = W_o \cdot \text{concat}(o_i^1, o_i^2, \ldots, o_i^H) \quad (3\text{-}8)$$

where H represents the number of attention heads.

Through this alignment, numerical patches are associated with textual prototypes, rendering numerical features in a natural language format suitable for LLM processing.

The prompt text $P$ is merged with the aligned patch embeddings $e_i'$, creating the final input sequence

$$I = [P, e_1', e_2', \ldots, e_k'] \quad (3\text{-}9)$$

### 2.4. LLM Inference

The sequence is fed into the frozen pre-trained LLM. The pre-trained LLM leverages its pattern recognition and inference abilities, developed from large-scale text training, to process the input, generating the output

$$O = \text{LLM}(I) \quad (3\text{-}10)$$

In this paper, Qwen2.5 is selected as the base LLM. It is a large-scale pre-trained LLM based on the Transformer architecture designed to enhance language understanding, reasoning capabilities, and multi-modal processing abilities. The model is pre-trained on a massive dataset of 18 trillion tokens, supporting 29 languages, including Chinese, English, French, and Spanish. Compared to previous versions, Qwen2.5 has achieved significant improvements in instruction following, long-text generation, structured data processing, and diverse system prompt responses. It performs exceptionally well in complex reasoning tasks, particularly in chain reasoning and tool-integrated reasoning tasks, demonstrating strong cross-modal adaptability. Qwen2.5 offers multiple parameter-scale versions, with the 3B model achieving a good balance between performance and computational efficiency. Despite having a smaller parameter size, Qwen2.5-3B demonstrates exceptional performance across benchmarks, achieving scores of 65.6, 42.1, and 42.6 in Massive Multitask Language Understanding (MMLU) [27], HumanEval [28], and MATH [29] evaluations, respectively, outperforming Llama3-8B, which scores 66.6, 20.5, and 33.5 on the same benchmarks. This indicates that Qwen2.5-3B has an efficient knowledge density and task adaptability, excelling in few-shot and zero-shot learning scenarios. This study employs the 3B version of Qwen2.5 as the new backbone of Time-LLM, integrating input reprogramming and prompt optimization to harness its strong sequence modeling abilities, offering reliable and efficient support for DPV power forecasting.

## 2.5. Output Projection

Using a flatten and linear projection layer $f_{proj}$, the LLM's high-dimensional output $O$ is transformed into the final forecasts. The flatten operation reshapes the multi-dimensional tensor $O$ into a lower-dimensional vector, preserving its spatial or temporal structure and enabling efficient processing. The linear projection layer $Linear_{proj}$ applies a learned weight matrix and bias vector to map the flattened vector to the target dimensionality of the forecast

$$forecast = f_{proj}(O) = Linear_{proj}(Flatten(O)) \tag{3-11}$$

In conclusion, Time-LLM employs FFT to extract periodic features, optimizes inputs using prompt texts, and leverages patch reprogramming and alignment mechanisms to convert numerical data into a format comprehensible to LLMs. High-dimensional representations are then generated using a frozen pre-trained LLM, and specific predictions are obtained through output projection. This method transforms time series forecasting into a natural language processing task, fully exploiting the strengths of LLMs in pattern recognition, contextual comprehension, inductive reasoning, and long-sequence modeling.

## 3. Experimental Results

### 3.1. Dataset and metrics

In this study, we selected datasets from California as a case study. These datasets are part of a nationwide dataset created by the National Renewable Energy Laboratory, generated using the sub-hour irradiance algorithm [30], covering the entire year of 2006 with a time resolution of 5 minutes. For this research, three DPV power plants were chosen from the California dataset for experimental analysis. Fig. 4 shows the geographical locations of the selected DPV power plants in California. The detailed characteristics of these power plants are summarized in Table 1.

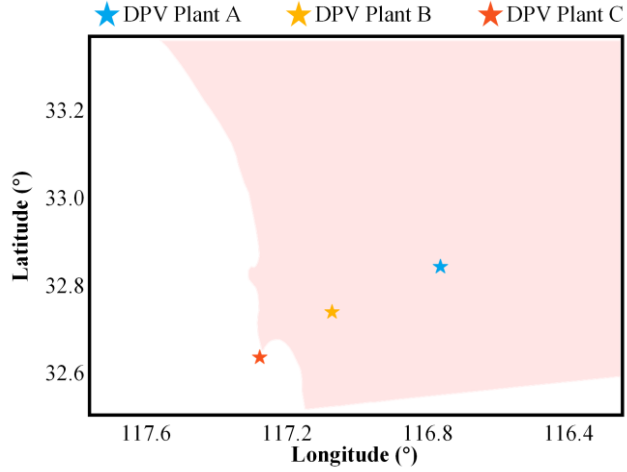

**Fig. 4. The locations of three DPV plants in southern California, marked with stars of different colors.**

Table 1 Summary of DPV Plants and Data Characteristics

| Plant | Coordinate | Time resolution | Installed Capacity |
|---|---|---|---|
| A | (117.25, 32.65) | 5 min | 13MW |
| B | (117.05, 32.75) | 5 min | 8MW |
| C | (116.75, 32.85) | 5 min | 8MW |

During the preprocessing of the dataset, we applied a normalization method by dividing the power by the installed capacity, enabling data from plants with different capacities to be compared on the same scale while reducing the impact of plant size on evaluation metrics. Additionally, we used cubic spline interpolation to handle missing and abnormal data, ensuring data completeness and consistency, thereby improving the accuracy of model training and forecasts.

To better align the dataset context in Time-LLM with the specifics of this study, we incorporated the following description: The distributed photovoltaic dataset includes output power measurements from distributed photovoltaic systems located in California, USA. It includes only photovoltaic power data and spans the entire year of 2006. Furthermore, the data exhibits a periodic pattern with no power generation during nighttime and early morning hours.

For the evaluation of long-term forecasting, we used mean absolute error (MAE), MSE, and the coefficient of determination ($R^2$) as performance metrics. MAE measures the mean absolute error between predicted values and ground truths, which is intuitive and less sensitive to outliers. MSE amplifies the impact of larger errors by squaring the differences, making it more sensitive to outliers. R² evaluates the goodness of fit of the model to the data, ranging from 0 to 1, with values closer to 1 indicating stronger explanatory power for data variance. For the evaluation of short-term forecasting models, we additionally employed symmetric mean absolute percentage error (SMAPE) [31] as a performance metric. SMAPE normalizes the absolute difference between predicted values and ground truths relative to their average value, ensuring error symmetry and making it suitable for evaluating datasets where predicted values and ground truths are of similar ranges. The formula is as follows

$$MAE = \frac{1}{N}\sum_{n=1}^{N}|Y_n - \hat{Y}_n| \tag{3-12}$$

$$MSE = \frac{1}{N}\sum_{n=1}^{N}(Y_n - \hat{Y}_n)^2 \tag{3-13}$$

$$R^2 = 1 - \frac{\sum_{n=1}^{N}(Y_n - \hat{Y}_n)^2}{\sum_{n=1}^{N}(Y_n - \overline{Y})^2} \tag{3-14}$$

$$SMAPE = \frac{200}{N}\sum_{n=1}^{N}\frac{|Y_n - \hat{Y}_n|}{|Y_n| + |\hat{Y}_n|} \tag{3-15}$$

where $N$ represents the total number of data points, corresponding to the forecast horizon, and $Y_n$ and $\hat{Y}_n$ denote the $n$-th ground truth and forecast, respectively, and $\overline{Y}$ represents the average of all ground truths.

These evaluation metrics offer a multidimensional view of model performance, enabling a thorough assessment of the model's capabilities in various tasks.

**3.2. Experiment setup and model training**

All experiments were performed in a high-performance computing setup with two NVIDIA H800 GPUs, an Intel(R) Xeon(R) Platinum 8468 CPU, and 486 GB of memory to meet computational demands.

During Time-LLM training, mean squared error (MSE) was used as the loss function, and the Adam [32] optimizer was employed for parameter updates. Only the lightweight modules, including the linear mapping layer, multi-head attention layer, and output projection layer, were trained, while the pre-trained LLM remained frozen. Similarly, baseline models were also trained using the Adam optimizer and MSE as the loss function to ensure a fair comparison. To ensure robust model training and reliable test outcomes, the dataset was partitioned into 70% for training, 20% for validation, and 10% for testing. To ensure the reliability of the results, each model was independently repeated three times in the experiments.

We compared the employed model with several advanced models in the field of time series forecasting. The selected baseline models were chosen based on their relevance and performance in capturing complex temporal dependencies and their diversity in architecture. The Transformer-based models PatchTST [33], Informer [34], and Autoformer [35] were included due to their strong performance in modeling long-range dependencies and handling various time series forecasting tasks recently, making them highly relevant for comparison. Additionally, the convolutional neural network-based model TimesNet [36] was selected as it offers advantages in handling multi-scale temporal features, which can be beneficial for capturing the dynamic behavior of DPV power generation. The competitive MLP-based model DLinear [37] was included for its simplicity and efficiency in handling time series data with a focus on leveraging linear dependencies, providing a contrasting approach to the more complex models. These models represent a broad range of contemporary advanced approaches, each with strengths in different aspects of time series forecasting, allowing for a comprehensive evaluation of Time-LLM's performance.

### 3.3. Short-term forecast

The time scale of short-term forecasting is usually small, typically at the hourly or minute level. In this study, we set two forecasting horizons, 12 and 24 (corresponding to one hour and two hours, respectively), to comprehensively evaluate the model's performance across different short-term forecasting tasks. To ensure the model effectively captures sufficient historical information for forecasting, the input sequence length was set to twice the forecasting horizon, providing the model with richer contextual information.

Experimental results, as shown in Table 2, indicate that our approach exhibits outstanding performance in short-term DPV power forecasting tasks. For forecast horizons of 12 and 24, the model surpasses other methods in MSE, MAE, $R^2$, and SMAPE metrics, highlighting its superior capacity to model short-term dynamics and intricate trends. Specifically, the model delivers over a 33% average reduction in MSE compared to the best conventional methods. In the SMAPE metric, it attains an improvement exceeding 1.5%. Regarding $R^2$, the model achieves an improvement of over 4%, demonstrating its ability to consistently provide accurate forecasts even for complex DPV power fluctuations.

Table 2 Short-term DPV power forecasting results. The forecasting horizons are 12 and 24 and the four rows provided are averaged from all three DPV plants. A lower value indicates better performance. Red: the best, Blue: the second best.

| Methods | Our | | TimesNet | | PatchTST | | DLinear | | Autoformer | | Informer | |
|---------|-----|-----|----------|-----|----------|-----|---------|-----|------------|-----|----------|-----|
| Horizon | 12 | 24 | 12 | 24 | 12 | 24 | 12 | 24 | 12 | 24 | 12 | 24 |
| MSE | 0.0031 | 0.0055 | 0.0595 | 0.1581 | 0.0045 | 0.0084 | 0.0046 | 0.0085 | 0.0106 | 0.0551 | 0.0280 | 0.0654 |
| MAE | 0.0346 | 0.0461 | 0.1893 | 0.3080 | 0.0325 | 0.0472 | 0.0413 | 0.0681 | 0.0589 | 0.1501 | 0.1334 | 0.1991 |
| $R^2$ | 0.9424 | 0.8991 | 0.1158 | 0 | 0.9172 | 0.8444 | 0.9153 | 0.8438 | 0.8060 | 0 | 0.4941 | 0 |
| SMAPE | 133.61 | 136.51 | 147.45 | 156.46 | 138.93 | 143.00 | 137.28 | 137.86 | 142.41 | 158.75 | 146.30 | 149.74 |

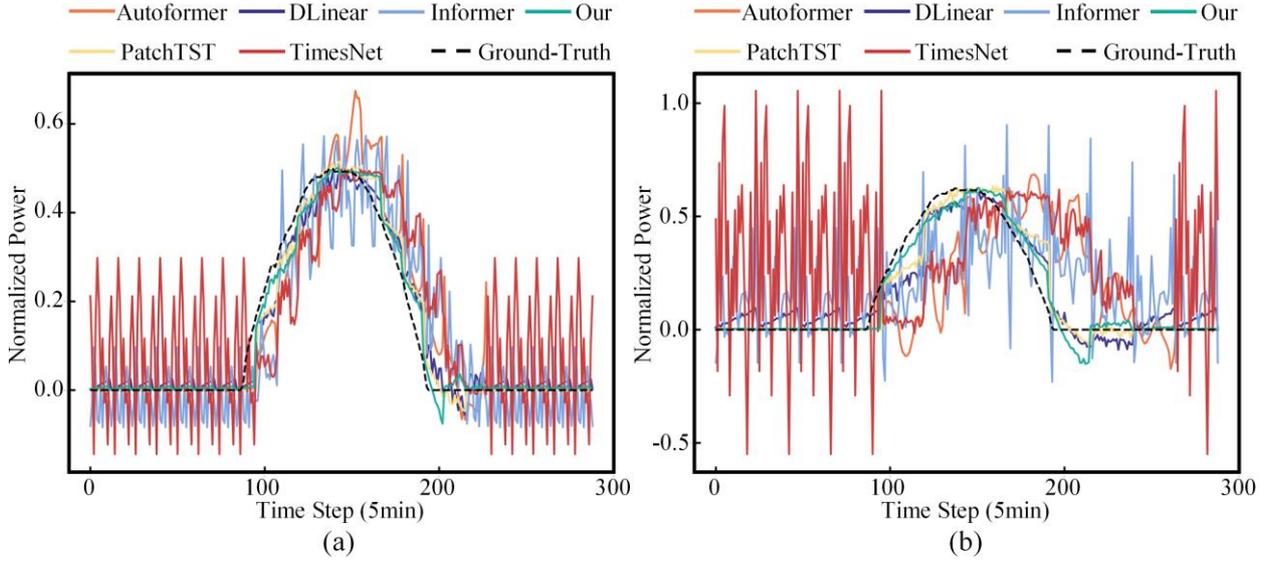

**Fig. 5 Short-term forecasting cases from the plant A by different models.** (a) under the input-24-forcast-12 settings. (b) under the input-48-forcast-24 settings.

The enhancement in performance stems not only from the model's precise modeling of local time series features but also its robust adaptability to sudden variations and noise. These capabilities are further validated by the visual comparisons in Fig. 5, which illustrate the short-term forecasting cases for plant A under two different experimental settings. Notably, competing methods such as Informer and Autoformer exhibit significant deviations from the ground-truth in certain periods, while our model maintains high accuracy and consistency. By capturing key patterns and features in short-term fluctuations, the model achieves greater stability and reliability in handling short-term forecasting tasks, laying a solid foundation for accurate forecast of DPV power dynamics.

### 3.4. Long-term forecast

In this study, we set the input time series length to 336, corresponds to a 28-hour period, to ensure the model can capture sufficiently long historical information, thereby improving forecast accuracy. To comprehensively evaluate the model's forecasting performance across different time scales, we selected two forecasting horizons, 192 and 336, which correspond to forecast intervals of 16 and 28 hours, respectively.

The experimental results, as shown in Table 3, confirm Time-LLM's exceptional performance in long-term forecasting tasks for DPV power forecast. In the experiments, Time-LLM showed significant superiority over existing best-performing models (e.g., PatchTST and DLinear) in MSE and $R^2$ metrics, especially for forecast horizons of 336. This exceptional performance stems from the well-designed input-output mapping mechanism in Time-LLM, which effectively mitigates sparsity and redundancy issues in time series data. By employing an effective sequence processing strategy, longer input sequences prevent information overload without noticeably increasing computational costs. Long input sequences enable a balance between capturing short-term fluctuations and long-term trends.

Table 3 Long-term DPV power forecasting results. A lower value indicates better performance. Red: the best, Blue: the second best.

| Plants | | A | | B | | C | |
|---|---|---|---|---|---|---|---|
| Horizon | | 192 | 336 | 192 | 336 | 192 | 336 |
| Our | MSE | 0.0053 | 0.0049 | 0.0055 | 0.0050 | 0.0045 | 0.0049 |
| | MAE | 0.0453 | 0.0392 | 0.0473 | 0.0390 | 0.0368 | 0.0411 |
| | $R^2$ | 0.8916 | 0.8996 | 0.9019 | 0.9097 | 0.9241 | 0.9181 |
| TimesNet | MSE | 0.0094 | 0.0096 | 0.0111 | 0.0113 | 0.0098 | 0.0094 |
| | MAE | 0.0598 | 0.0602 | 0.0621 | 0.0631 | 0.0598 | 0.0576 |
| | $R^2$ | 0.8061 | 0.8018 | 0.8010 | 0.7973 | 0.8356 | 0.8425 |
| PatchTST | MSE | 0.0065 | 0.0065 | 0.0070 | 0.0069 | 0.0066 | 0.0068 |
| | MAE | 0.0475 | 0.0490 | 0.0484 | 0.0480 | 0.0495 | 0.0492 |
| | $R^2$ | 0.8674 | 0.8666 | 0.8734 | 0.8763 | 0.8899 | 0.8862 |
| DLinear | MSE | 0.0075 | 0.0075 | 0.0076 | 0.0073 | 0.0075 | 0.0073 |
| | MAE | 0.0510 | 0.0516 | 0.0496 | 0.0485 | 0.0500 | 0.0485 |
| | $R^2$ | 0.8458 | 0.8454 | 0.8627 | 0.8686 | 0.8750 | 0.8772 |
| Autoformer | MSE | 0.0601 | 0.0516 | 0.0663 | 0.0594 | 0.0660 | 0.0595 |
| | MAE | 0.2118 | 0.1931 | 0.2242 | 0.2056 | 0.2275 | 0.2098 |
| | $R^2$ | 0 | 0 | 0 | 0 | 0 | 0.0020 |
| Informer | MSE | 0.0093 | 0.0099 | 0.0103 | 0.0112 | 0.0103 | 0.0116 |
| | MAE | 0.0654 | 0.0701 | 0.0669 | 0.0731 | 0.0656 | 0.0729 |
| | $R^2$ | 0.8079 | 0.7955 | 0.8138 | 0.7978 | 0.8279 | 0.8051 |

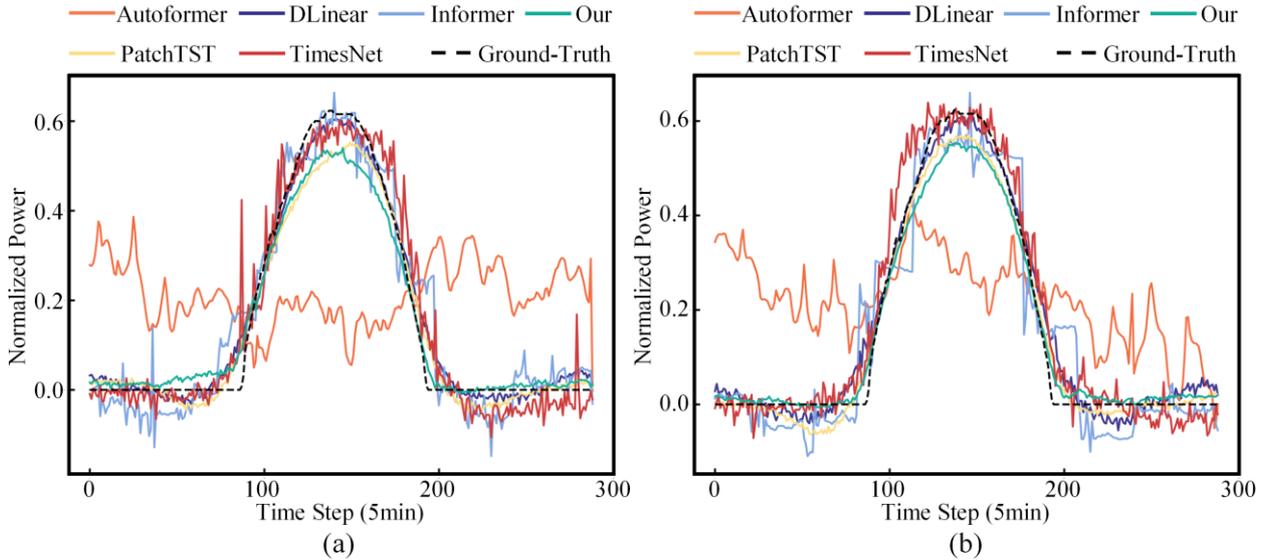

**Fig. 6 Long-term forecasting cases from the plant A by different models.** (a) under the input-336-forcast-192 settings. (b) under the input-336-forcast-336 settings.

The advantages of Time-LLM are further illustrated in Fig. 6, revealing that Time-LLM (Our) aligns better with the ground-truth power outputs across various power levels, including sharp transitions and zero-power zones.

Compared with PatchTST, Time-LLM delivers over 25% and 14% average reductions in MSE and MAE, respectively, along with more than 3.5% improvement in $R^2$. The results clearly demonstrate that Time-LLM excels in both short-term and long-term forecasting, offering substantial advantages in complex dynamic modeling for DPV power forecast.

## 3.5. Few-shot forecast

In recent years, LLMs have demonstrated outstanding performance in few-shot learning tasks [38], showcasing their strong generalization ability under limited data conditions. This section evaluates the few-shot learning ability of our model in forecasting tasks. To ensure fairness in evaluation, we followed the experimental setup from relevant studies and tested under constrained training data conditions, limiting the training data to 50%, 20%, 10%, and 5% of the original dataset.

Experimental results, as shown in Table 4 and Fig. 7, indicate that Time-LLM exhibits significant advantages in few-shot learning scenarios, particularly under conditions with low training data proportions (e.g., 5% and 10%). Across metrics such as MSE, MAE, and $R^2$, Time-LLM consistently surpasses other baseline models in most cases. This indicates that Time-LLM is capable of effectively capturing time series patterns and delivering accurate forecasts even under data-limited circumstances. This performance may be attributed to Time-LLM's ability to successfully activate the knowledge embedded in pre-trained LLMs, enabling it to uncover latent patterns in time series with limited training data. For the $R^2$ metric, Time-LLM's superiority becomes especially evident under low data proportion scenarios, delivering forecasts that markedly excel in interpretability and consistency compared to other models. The $R^2$ of Time-LLM remains consistently high, whereas other baseline models (such as Autoformer) display notable instability and even underfitting challenges under similar conditions. This further demonstrates Time-LLM's adaptability and robustness in few-shot learning scenarios.

Table 4 Few-shot learning on different training data. All results are averaged from four different DPV plants: P ∈ {A, B, C}. Red: the best, Blue: the second best.

| Limitation | | 5% | | 10% | | 20% | | 50% | |
|---|---|---|---|---|---|---|---|---|---|
| Horizon | | 192 | 336 | 192 | 336 | 192 | 336 | 192 | 336 |
| Our | MSE | 0.0062 | 0.0061 | 0.0062 | 0.0057 | 0.0056 | 0.0055 | 0.0065 | 0.0055 |
| | MAE | 0.0457 | 0.0455 | 0.0474 | 0.0433 | 0.0462 | 0.0437 | 0.0499 | 0.0452 |
| | $R^2$ | 0.8861 | 0.8882 | 0.8841 | 0.8942 | 0.8969 | 0.8985 | 0.8788 | 0.8977 |
| TimesNet | MSE | 0.0223 | 0.0144 | 0.0151 | 0.0142 | 0.0132 | 0.0109 | 0.0148 | 0.0102 |
| | MAE | 0.0904 | 0.0686 | 0.0729 | 0.0721 | 0.0672 | 0.0600 | 0.0727 | 0.0599 |
| | $R^2$ | 0.5919 | 0.7372 | 0.7225 | 0.7381 | 0.7563 | 0.7992 | 0.7278 | 0.8119 |
| PatchTST | MSE | 0.0067 | 0.0065 | 0.0083 | 0.0082 | 0.0072 | 0.0070 | 0.0070 | 0.0070 |
| | MAE | 0.0477 | 0.0480 | 0.0520 | 0.0546 | 0.0488 | 0.0485 | 0.0512 | 0.0506 |
| | $R^2$ | 0.8766 | 0.8789 | 0.8464 | 0.8468 | 0.8683 | 0.8728 | 0.8715 | 0.8720 |
| DLinear | MSE | 0.0076 | 0.0073 | 0.0079 | 0.0076 | 0.0071 | 0.0068 | 0.0080 | 0.0076 |
| | MAE | 0.0502 | 0.0483 | 0.0505 | 0.0490 | 0.0449 | 0.0436 | 0.0550 | 0.0524 |
| | $R^2$ | 0.8592 | 0.8651 | 0.8536 | 0.8594 | 0.8698 | 0.8741 | 0.8519 | 0.8590 |
| Autoformer | MSE | 0.0645 | 0.0592 | 0.0644 | 0.0598 | 0.0660 | 0.0574 | 0.0626 | 0.0566 |
| | MAE | 0.2224 | 0.2096 | 0.2210 | 0.2073 | 0.2227 | 0.2039 | 0.2200 | 0.2037 |
| | $R^2$ | 0 | 0 | 0 | 0 | 0 | 0 | 0 | 0 |
| Informer | MSE | 0.0115 | 0.0108 | 0.0098 | 0.0112 | 0.0099 | 0.0107 | 0.0096 | 0.0109 |
| | MAE | 0.0810 | 0.0735 | 0.0685 | 0.0735 | 0.0689 | 0.0709 | 0.0667 | 0.0721 |
| | $R^2$ | 0.7909 | 0.8037 | 0.8214 | 0.7952 | 0.8182 | 0.8025 | 0.8227 | 0.8002 |

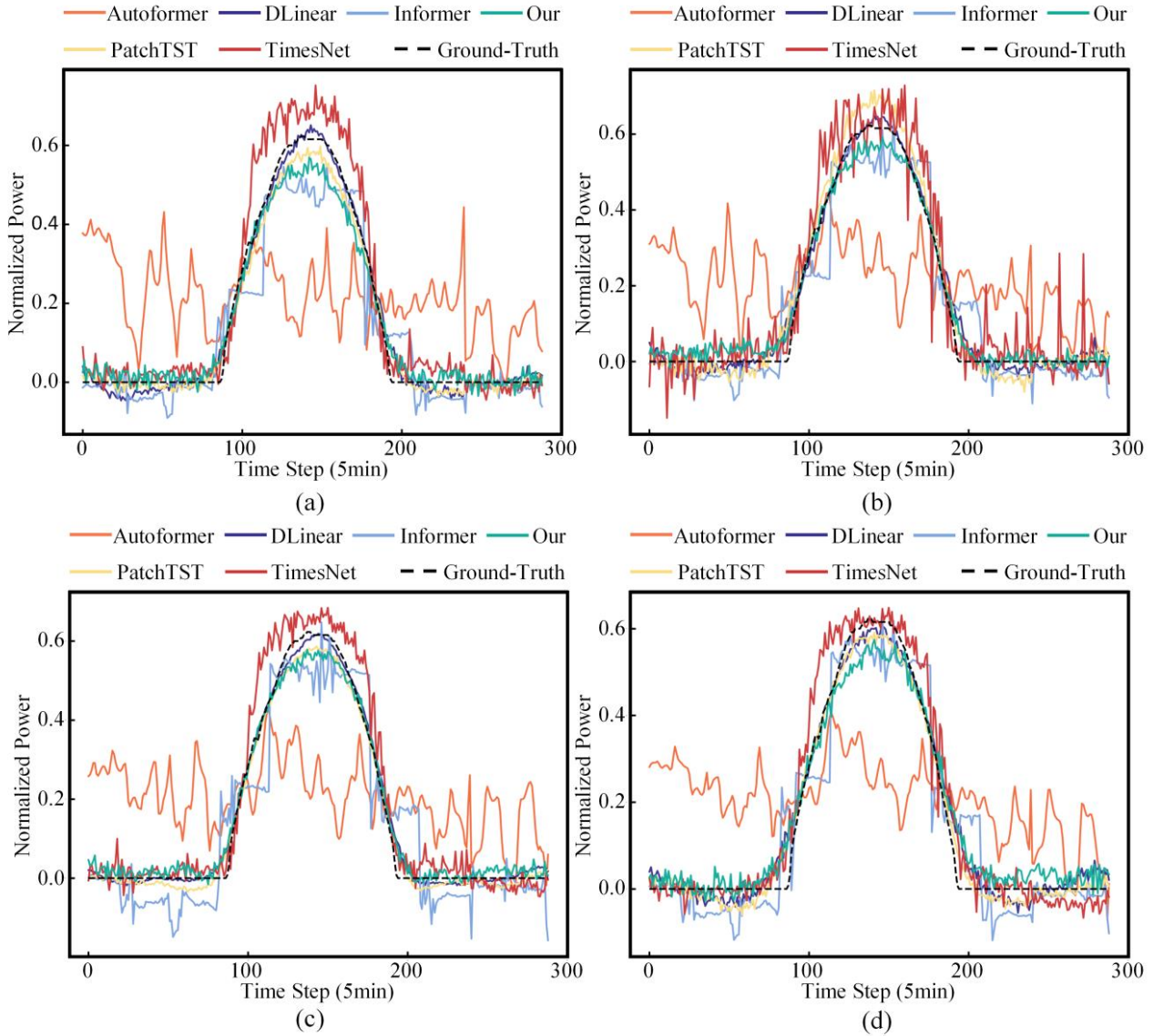

**Fig. 7 Few-shot forecasting cases from plant A by different models.** (a) under the 5% training dataset and input-336-forcast-336 settings. (b) under the 10% training dataset and input-336-forcast-336 settings. (c) under the 20% training dataset and input-336-forcast-336 settings. (d) under the 50% training dataset and input-336-forcast-336 settings.

### 3.6. Zero-shot forecast

Beyond their few-shot learning capabilities, LLMs demonstrate remarkable potential in efficient zero-shot inference. In this study, we evaluate the zero-shot learning and transfer capabilities of Time-LLM through a cross-plant adaptation framework, using training data from one plant to predict data from another unseen plant.

In traditional DPV power forecasting approaches, DPV systems often necessitate separate training for every new plant. This method greatly escalates computational resource requirements and prolongs deployment timelines, thereby constraining its practical implementation in large-scale distributed systems. In contrast, Time-LLM leverages the strong generalization capabilities of pre-trained LLMs to achieve efficient zero-shot forecast without requiring additional training for new plants.

Compared to Time-LLM, other baseline models perform inadequately in cross-plant tasks, with limited generalization capabilities in cross-plant transfer tasks.

Experimental results, as shown in Table 5 and Fig. 8, indicate that Time-LLM exhibits notable performance advantages in cross-plant tasks. For the MSE metric, Time-LLM reduces forecast errors by an average of 10.7% compared to PatchTST and by 18% relative to DLinear, substantially minimizing forecast errors. In terms of MAE, Time-LLM achieves 10.3% and 10.7% higher predictive accuracy than PatchTST and DLinear, respectively, showcasing enhanced stability and reliability. Regarding the $R^2$ metric, Time-LLM improves predictive consistency by 1.5% compared to PatchTST and by 3% compared to DLinear, fully demonstrating its ability to capture complex time series patterns. Overall, these experimental results not only validate the efficiency of Time-LLM in cross-plant tasks but also further reinforce its significant advantages in generalization and robustness.

Although these models excel in single task, their customized nature limits their generalization ability, making it difficult for them to adapt to other tasks and scenarios. In contrast, Time-LLM employs reprogramming and prompt mechanisms to directly utilize the language understanding capabilities of pre-trained LLMs for time series data processing, without requiring modifications to the model structure or extensive training data. This method not only greatly boosts the model's generalization and scalability but also strengthens its adaptability in cross-plant tasks, showcasing the potential of LLMs in multi-task and multi-domain time series forecasting.

Table 5 Zero-shot learning results. <span style="color:red">Red</span>: the best, <span style="color:blue">Blue:</span> the second best.

| Dataset | | A->B | | A->C | | B->A | | B->C | | C->A | | C->B | |
|---|---|---|---|---|---|---|---|---|---|---|---|---|---|
| Horizon | | 192 | 336 | 192 | 336 | 192 | 336 | 192 | 336 | 192 | 336 | 192 | 336 |
| Our | MSE | 0.0064 | 0.0063 | 0.0066 | 0.0059 | 0.0060 | 0.0054 | 0.0063 | 0.0063 | 0.0051 | 0.0071 | 0.0058 | 0.0062 |
| | MAE | 0.0491 | 0.0442 | 0.0503 | 0.0428 | 0.0506 | 0.0411 | 0.0519 | 0.0437 | 0.0390 | 0.0481 | 0.0398 | 0.0454 |
| | $R^2$ | 0.8841 | 0.8866 | 0.8895 | 0.9014 | 0.8763 | 0.8896 | 0.8943 | 0.8949 | 0.8949 | 0.8536 | 0.8962 | 0.8885 |
| TimesNet | MSE | 0.0099 | 0.0098 | 0.0094 | 0.0097 | 0.0105 | 0.0108 | 0.0106 | 0.0111 | 0.0098 | 0.0092 | 0.0102 | 0.0095 |
| | MAE | 0.0592 | 0.0593 | 0.0579 | 0.0588 | 0.0624 | 0.0633 | 0.0616 | 0.0625 | 0.0615 | 0.0591 | 0.0609 | 0.0582 |
| | $R^2$ | 0.8220 | 0.8235 | 0.8425 | 0.8365 | 0.7849 | 0.7787 | 0.8215 | 0.8132 | 0.7992 | 0.8100 | 0.8164 | 0.8291 |
| PatchTST | MSE | 0.0068 | 0.0066 | 0.0064 | 0.0067 | 0.0068 | 0.0067 | 0.0068 | 0.0068 | 0.0068 | 0.0067 | 0.0070 | 0.0068 |
| | MAE | 0.0472 | 0.0484 | 0.0464 | 0.0490 | 0.0489 | 0.0481 | 0.0476 | 0.0479 | 0.0510 | 0.0492 | 0.0505 | 0.0491 |
| | $R^2$ | 0.8785 | 0.8810 | 0.8927 | 0.8878 | 0.8605 | 0.8626 | 0.8861 | 0.8856 | 0.8603 | 0.8632 | 0.8747 | 0.8770 |
| DLinear | MSE | 0.0076 | 0.0074 | 0.0073 | 0.0074 | 0.0076 | 0.0074 | 0.0073 | 0.0073 | 0.0077 | 0.0075 | 0.0078 | 0.0074 |
| | MAE | 0.0494 | 0.0498 | 0.0489 | 0.0495 | 0.0516 | 0.0511 | 0.0486 | 0.0479 | 0.0531 | 0.0519 | 0.0511 | 0.0493 |
| | $R^2$ | 0.8636 | 0.8663 | 0.8774 | 0.8754 | 0.8447 | 0.8471 | 0.8777 | 0.8781 | 0.8416 | 0.8454 | 0.8599 | 0.8674 |
| Autoformer | MSE | 0.0643 | 0.0586 | 0.0697 | 0.0606 | 0.0624 | 0.0523 | 0.0713 | 0.0602 | 0.0569 | 0.0504 | 0.0610 | 0.0568 |
| | MAE | 0.2219 | 0.2060 | 0.2321 | 0.2111 | 0.2147 | 0.1929 | 0.2338 | 0.2094 | 0.2075 | 0.1917 | 0.2173 | 0.2035 |
| | $R^2$ | 0 | 0 | 0 | 0 | 0 | 0 | 0 | 0 | 0 | 0 | 0 | 0 |
| Informer | MSE | 0.0096 | 0.0107 | 0.0095 | 0.0110 | 0.0102 | 0.0107 | 0.0100 | 0.0114 | 0.0107 | 0.0112 | 0.0107 | 0.0116 |
| | MAE | 0.0648 | 0.0713 | 0.0647 | 0.0718 | 0.0684 | 0.0727 | 0.0661 | 0.0728 | 0.0688 | 0.0734 | 0.0670 | 0.0735 |
| | $R^2$ | 0.8268 | 0.8072 | 0.8410 | 0.8151 | 0.7899 | 0.7803 | 0.8320 | 0.8092 | 0.7809 | 0.7697 | 0.8068 | 0.7906 |

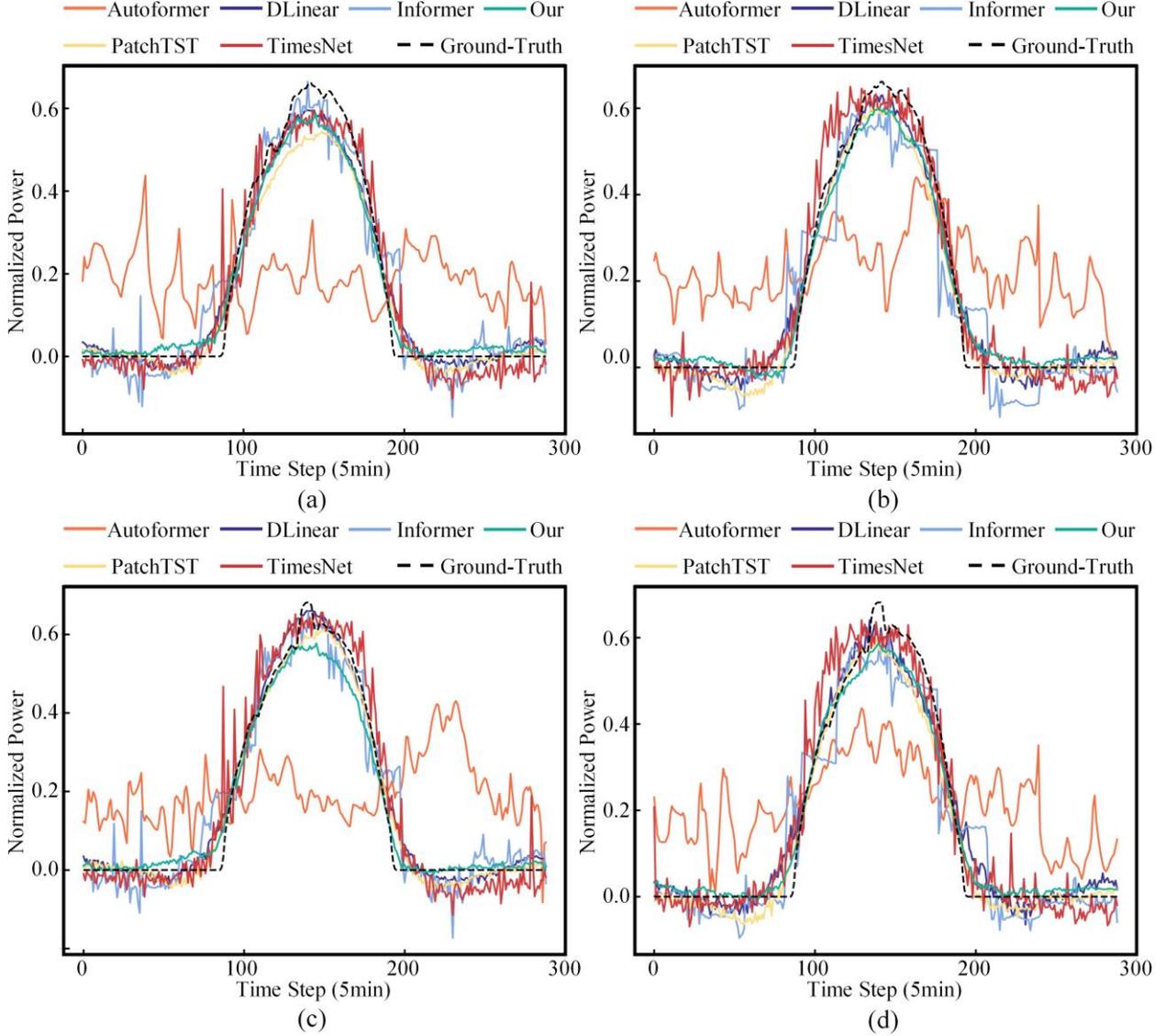

**Fig. 8 Zero-shot forecasting cases from Plant A by different models.** (a) from plant A -> plant B under the input-336-forcast-192 and the input-336-forcast-336 settings. (b) from plant A -> plant C under the input-336-forcast-192 and the input-336-forcast-336 settings.

## 4. Conclusion

This study systematically investigates the complex task of DPV power forecasting based on the Time-LLM framework, which aims to overcome the reliance of conventional approaches on high-quality external data and costly hardware and seeks to improve the model's utility and deployment feasibility through the exploitation of historical time series data. Experimental results reveal that the Time-LLM framework consistently delivers outstanding performance across diverse task scenarios. In short-term forecasting, the model accurately captures the dynamic changes of DPV power, with minimal error increase even when the forecasting range is significantly expanded, demonstrating good stability and robustness. Particularly in scenarios with rapid DPV power fluctuations, Time-LLM effectively captures critical pattern changes, highlighting its strong time series modeling capability. In long-term forecasting tasks, Time-LLM accurately models long-term dependencies, effectively adapting to the complexities of dynamic time series, which markedly enhances forecasting precision and consistency. In few-shot learning tasks, Time-LLM exhibits exceptional generalization, effectively discerning latent patterns from source plants and successfully adapting to target plant tasks with minimal training data, thereby substantially reducing data requirements. Furthermore, in zero-shot learning tasks, Time-LLM achieves effective adaptation relying solely on source plant data, eliminating the need for target plant training data, thereby demonstrating its versatility and adaptability in cross-plant tasks, particularly for DPV settings where data is difficult to obtain.

Future research could focus on developing lightweight models suitable for low-power devices to meet real-time forecasting requirements in edge computing scenarios. Additionally, uncertainty modeling methods for time series data could be further explored, enhancing the model's ability to represent future uncertainties through probabilistic forecasting or generative modeling techniques, thereby improving forecasting reliability and robustness.

**Acknowledgments**

This work was supported by the National Natural Science Foundation of China under Grant No. 52077194.

**CRediT authorship contribution statement**

**Huapeng Lin:** Conceptualization, Methodology, Software, Formal analysis, Visualization, Validation, Investigation, Writing- Original Draft. **Miao Yu:** Conceptualization, Methodology, Writing- Review & Editing, Supervision.

**Declaration of competing interest**

The data that support the findings of this study are available from the authors on reasonable request.

**Competing interests**

The authors declare that they have no known competing financial interests or personal relationships that could have appeared to influence the work reported in this paper.